\documentclass[twocolumn,showpacs,preprintnumbers,amsmath,amssymb]{revtex4-1}

\usepackage{graphicx}
\usepackage{amsmath}
\usepackage{amssymb}
\usepackage{bm}
\usepackage{tabularx}

\begin{document}


\title{Impulse response of the Bayreuth Festspielhaus}

\author{Kai Huang}
\email{kai.huang@uni-bayreuth.de}
\affiliation{Experimentalphysik V, Universit\"at Bayreuth, 95440 Bayreuth, Germany}

\date{\today}

\begin{abstract}
The Bayreuth Festspielhaus is well known for its architecture because its design is heavily influenced by composer Richard Wagner. Due to the special acoustic design, the reverberation time (i.e., time scale for the sound pressure level to decay $60$~dB) is larger than usual opera houses. Using hand-claps and smart phone recordings, I measured the impulse response of the Bayreuth Festspielhaus in the auditorium, on the stage, as well as in the orchestra pit. The measured reverberation time shows quantitative agreement with the literature values within a certain frequency range, demonstrating the possibility of using this approach to monitor room acoustics.
\end{abstract}


\maketitle
\section*{Introduction}
\label{sec:intro} 

The Bayreuth Festspielhaus is unique for its special acoustic design by Richard Wagner in the 19th century \cite{Beranek2010}. It has also been well preserved since then for Wagnerian operas. The hidden orchestra pit, for instance, clears the way for the audience to focus on the stage. Together with the reflecting board that blocks the direct sound, the pit acts as a low-pass filter and sets the sound from the orchestra in perspective with the singers \cite{Polack2011}. Due to the special design, the reverberation time (RT), which is one of the most important room acoustic quantities, of the Bayreuth Festspielhaus is relatively large compared with the other opera houses of the same age \cite{Beranek2010}. 

Impulse response is often used to characterize room acoustics, because it probes all frequency components of a room. Based on the measured response, further characterizations on the strength, speech intelligibility, echoes, reverberation, and other features are straightforward to be quantified \cite{Kuttruff2000,Mueller2012}. Such a quantification facilitates room acoustics design and planning, so that incommunicable rooms such as the one that Sabine was facing more than 100 years ago \cite{Sabine1922} can be avoided. The source of an impulse can be a loudspeaker fed with a pulse signal, exploding objects such as air-balloons, or hand-claps. Following a standard protocol \cite{iso}, we can determine RT and other room acoustic quantities from the recorded signals \cite{Kuttruff2000,Mueller2012}.  The RT and other room acoustic properties of the Bayreuth Festspielhaus have been measured systematically for more than half a century \cite{Beranek2010}. Shortly before the recent renovation, Garai and colleagues have conducted a systematic characterization of the Festspielhaus in 2014 \cite{Garai2015}.

With the recent development of information technology, room acoustic characterizations are becoming more convenient using hand claps and the microphones from smart phones \cite{Seetharaman2012,Rosenkranz2017}. An interesting follow-up question is: How well can the acoustics of a room be characterized using this approach in comparison to the conventional measuring protocol? Here, the impulse response of the Bayreuth Festspielhaus acquired by hand-claps and smart phone recordings (HCSP) is presented. The corresponding room acoustic parameters (RT and center time) are  obtained with the ITA-Toolbox \cite{Dietrich2013} and compared with literature values, which were collected in the same month as the present investigation (September, 2014) \cite{Garai2015}. Based on the comparison, I discuss the advantages and typical drawbacks of the HCSP and provide a checklist for collecting sensible data from this approach.   

\section*{Measuring Procedure}
\label{sec:methods}

\begin{figure}
    \begin{center}
		\includegraphics[width=0.9\columnwidth]{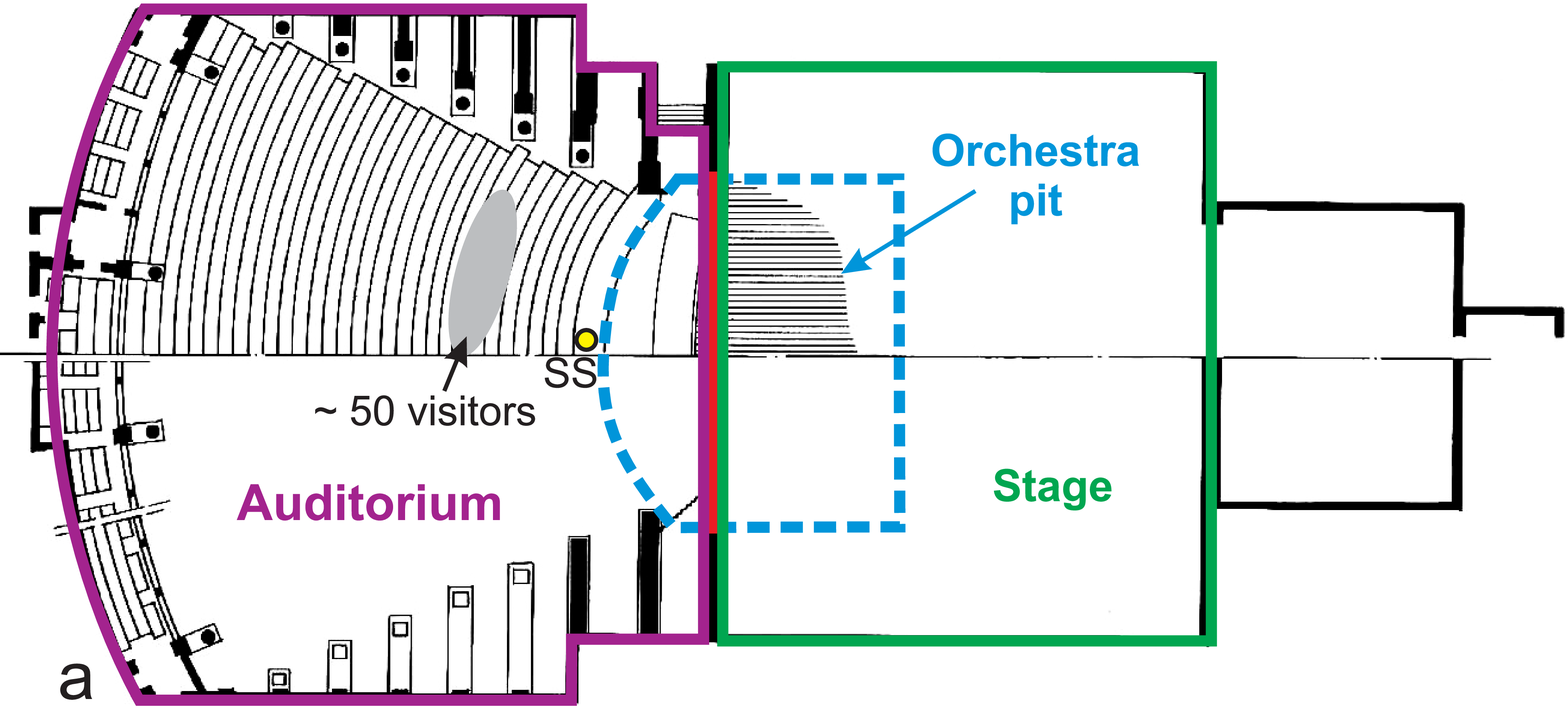} \\
		\includegraphics[width=0.8\columnwidth]{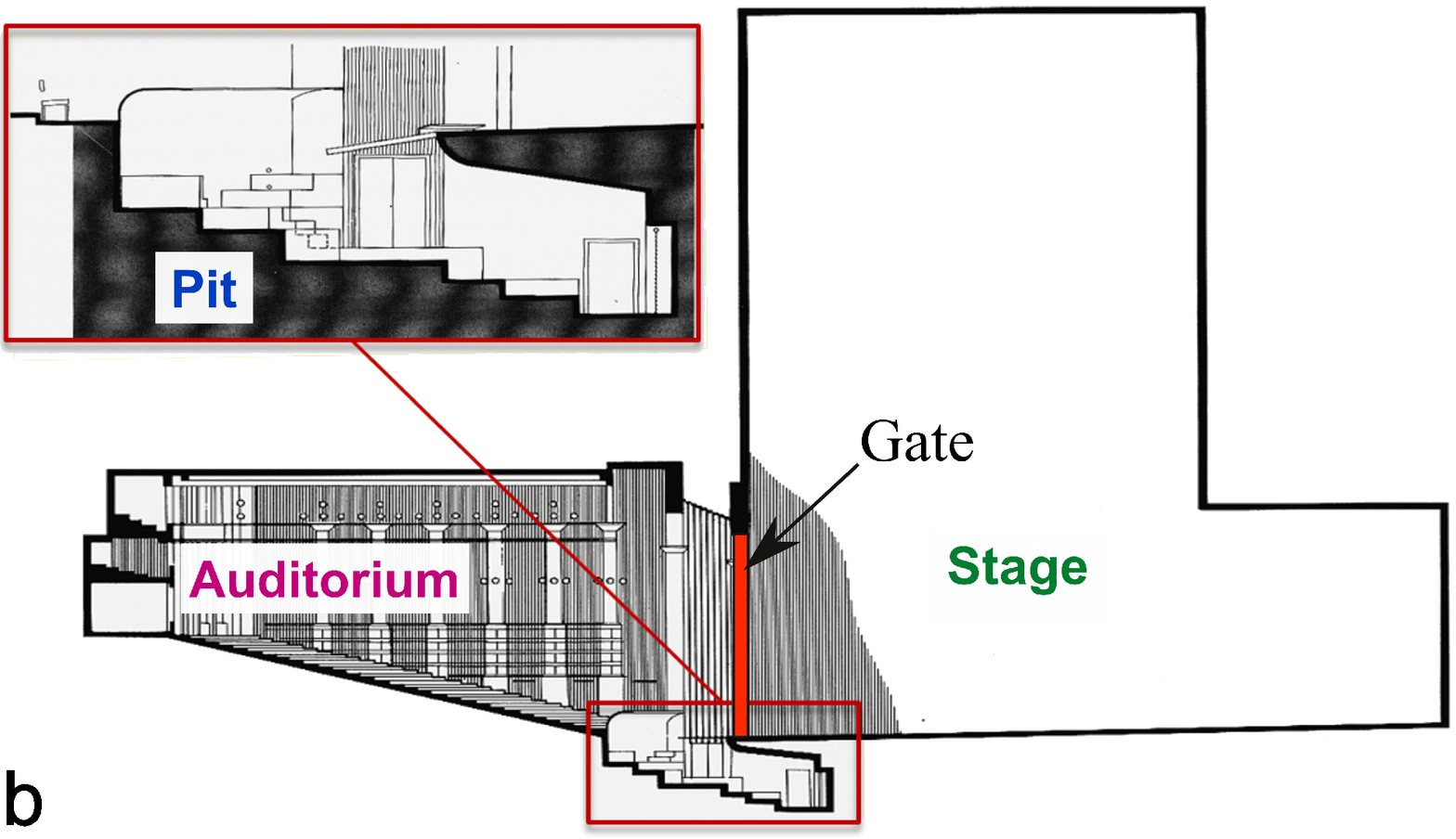}
    \end{center}
    \caption{Architects' drawings of the Bayreuth Festspielhaus adapted from Ref.~\cite{Beranek2010}. (a) A floor plan with the borders of the auditorium, the stage and the orchestra pit, the location of the sound source (SS) marked. The recording devices are located in the region shaded in gray. The red line represents the stage gate, which was closed during the measurement. (b) Corresponding side-view sketch with a close-view of the orchestra pit. }
    \label{fig:method}  
\end{figure}

Figure \ref{fig:method} shows the sketches of the Festspielhaus adapted from the architects' drawings. The measurement was conducted during a visit with about 50 visitors on September 11, 2014. As marked in Fig.\,\ref{fig:method}(a), the sound of five hand-claps was generated at location SS, directly in front of the orchestra pit. The sound was recorded by the microphones of three smart phones with a sampling rate of $44.1$~kHz. The distances between the source and recorders were $5\sim 10$ meters, roughly double the reverberation distance \cite{Mueller2012}. In addition, five hand-claps with both SS and microphones in the orchestra pit (with shorter distances between SS and the recorders due to the limited space) and on the stage were also performed. Subsequently, the recorded signal of each individual clap was extracted manually for post processing. 

The data analysis was performed with Matlab using the ITA-Toolbox \cite{Dietrich2013}, which provides a standard routine for room acoustic characterizations following ISO 3382 \cite{iso}. More specifically, the raw signals are filtered into different octave bands and subsequently the energy decay curve (EDC) for each band is obtained through applying the backward integration method on the squared envelope of each filtered signal \cite{Kuttruff2000}. From least square fits of the EDC in the semi-logarithmic plane, the reverberation time is obtained. Depending on the different ranges of data used in the fitting, EDT (early reverberation time), T15, T20, T30 (i.e., using the energy decay from $-5$~dB to $-20$, $-25$, or $-35$~dB), etc., are obtained. In the present study, T15 and the center time obtained in the frequency range from $250$~Hz to $4000$~Hz, covering five octave bands, are presented. T20 and T30 are not used because they cannot be obtained for all frequencies due to the limited signal-to-noise ratio (SNR).

\section*{Results and Discussion}
\label{sec:ir}

\begin{figure}[hbt]
    \begin{center}
        \includegraphics[width=0.9\columnwidth]{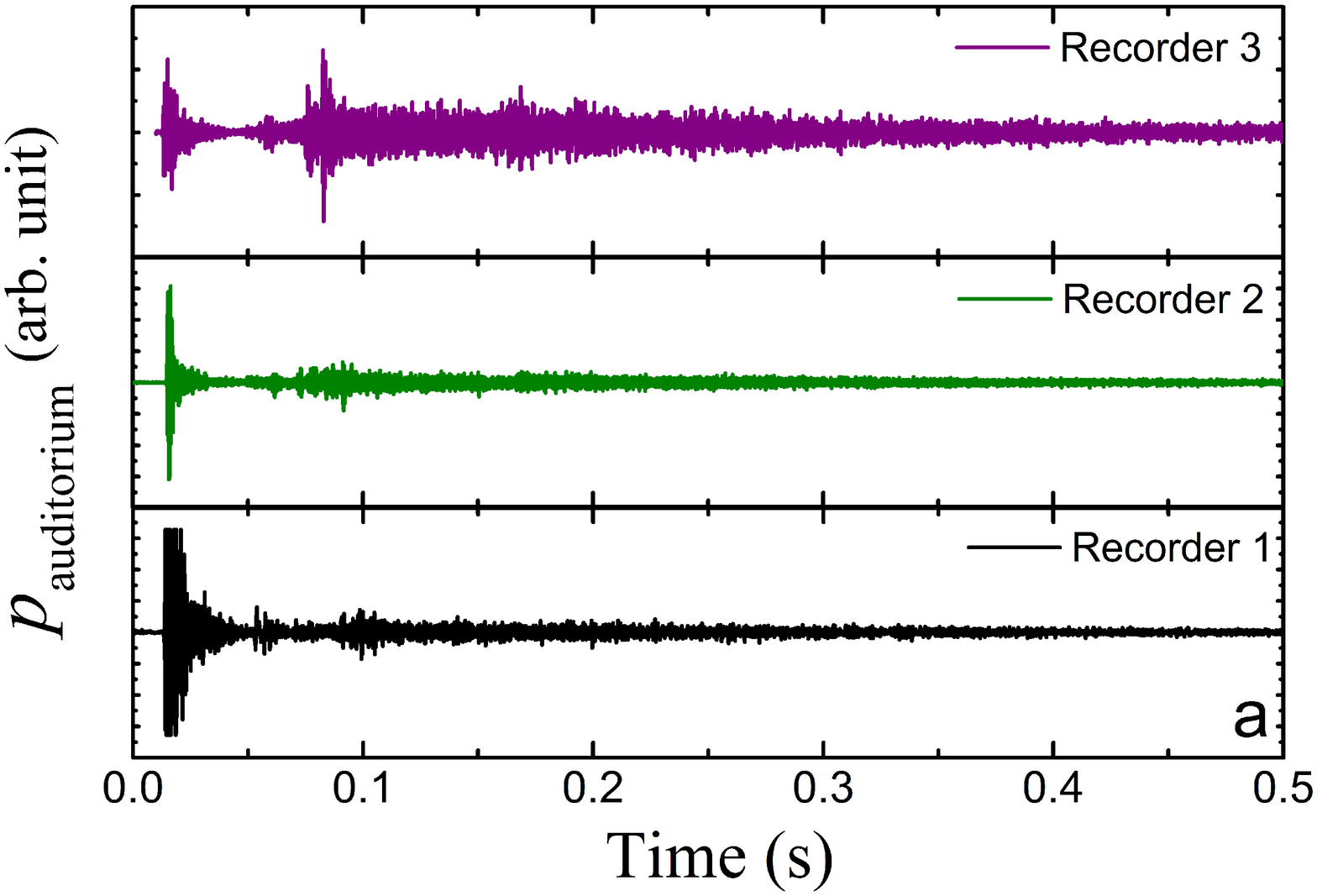} \\
		\vskip 1em
        \includegraphics[width=0.85\columnwidth]{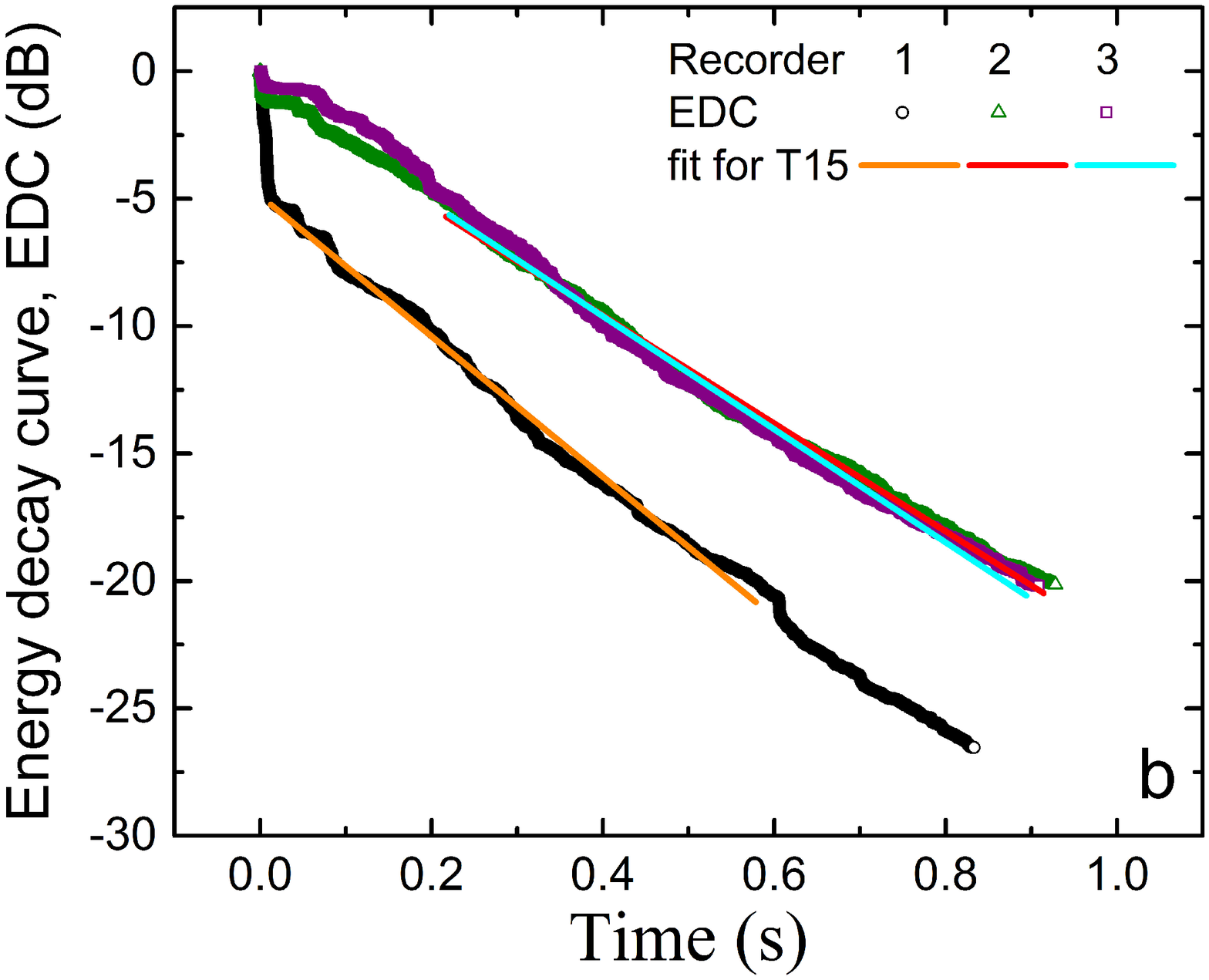} 
    \end{center}
    \caption{(a) Representative raw data captured by the three recorders in the Auditorium. (b) Corresponding energy decay curves for center frequency $1000$~Hz. The solid lines in (b) are linear fits of the data in the semi-logarithmic plane to obtain T15.}
    \label{fig:ir}
\end{figure}

Figure \ref{fig:ir}(a) shows the raw signals captured by the three microphones. Although all the smart phones use the same program (Smart Voice Recorder) for recording, the sound pressure differs one from another, owing to the different properties of the microphones and analogue-digital (AD) converters embedded, as well as the different distances to the SS. The following features can be learned from a comparison of the raw signals: (i) The direct and reflected sound can be clearly distinguished (separated at $\sim0.05$~s) for all recorded signals. (ii) the sound pressure of the direct sound may vary dramatically from one recording device to another. For instance, the maximum sound pressure from recorder 1 is more than one order of magnitude larger than that from recorder 3. (iii) The direct sound is not always the strongest peak in the impulse response. The difference of the direct sound energy will lead to the scattering of the center time, which will be discussed at the end of this section. 

The cumulative energy decay curves obtained with the backward integration method \cite{Schroeder1965} are shown in Fig.\,\ref{fig:ir}. For all three recordings, EDCs have a similar behavior: A stepwise initial drop followed by an exponential decay (note the semi-logarithmic scale). The initial drop of the energy level represents the influence of the direct sound. Because of the strong direct sound from recorder 1, the magnitude of the corresponding EDC drop, which represents the energy level of the direct sound, is much larger than the other two. Based on definition, RT is obtained through a fit of the exponential decay part of EDC. The agreement of the EDCs in this regime indicates that the exact locations and gain levels of the microphones play a minor role in obtaining RT, because those configurations only lead to different factors to the EDC, not the decay exponent. Because the sound energy generated by hand-claps is limited and the waiting time between subsequent claps is not sufficiently long, the conventional T30 cannot be obtained for all octave bands. Instead, T15 is used here. This difference can lead to additional uncertainty in determining RT. Moreover, as indicated by the EDC of recorder 1, too large direct sound energy may lead to a rapid drop of the energy level to below $-5$~dB and consequently a shorter reverberation time.

\begin{figure}[hbt]
    \begin{center}
        \includegraphics[width=0.85\columnwidth]{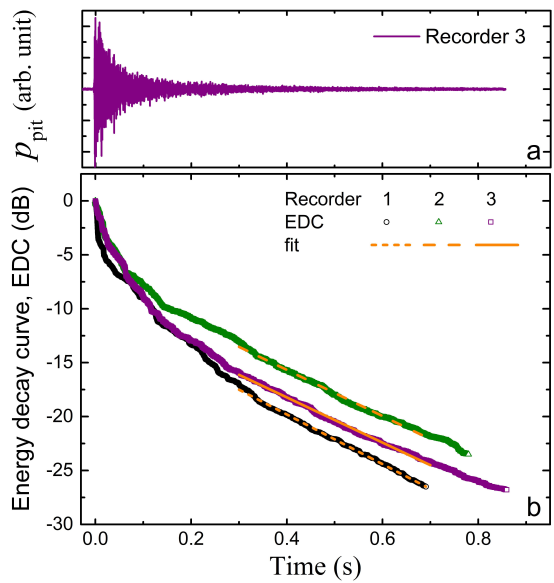}
    \end{center}
    \caption{(a) A sample recording obtained in the orchestra pit, illustrating the lack of clear separation between direct and reflected sound. (b) Corresponding EDCs at $1$~kHz for all three recorders. Solid, dashed and short dashed lines correspond to the linear fits of the data in the semi-logarithmic plane, using the energy decay between $0.30$ and $0.70$~s.}
    \label{fig:irp}
\end{figure}

As shown in Fig.\,\ref{fig:irp}(a), no clear distinguish between the direct and reflected sound can be made for the data collected in the orchestra pit. This can be attributed to the relatively small volume and the complex geometry of the pit [see the inset of Fig.\,\ref{fig:method}(b)]. In addition, the sound reflecting board, which was designed to block the direct sound to the stage, effectively mixes the directed and reflected sound together. Last but not least, the opening to the stage gives rise to additional influence from the stage and auditorium. Consequently, the EDCs shown in Fig.\,\ref{fig:irp}(b) exhibit a gradual decay instead of the rapid drop. Thus, using the conventional ways of determining RT is inappropriate, because the exponential decay starts at an energy level lower than $-5$~dB. Thus, I fit the tails of the EDCs and obtain RT from the slopes of the fits. An average of the three EDCs yields an RT of $1.38\pm0.08$~s for the octave band centered at $1000$~Hz. 

\begin{figure}[hbt]
    \begin{center}
        \includegraphics[width=0.8\columnwidth]{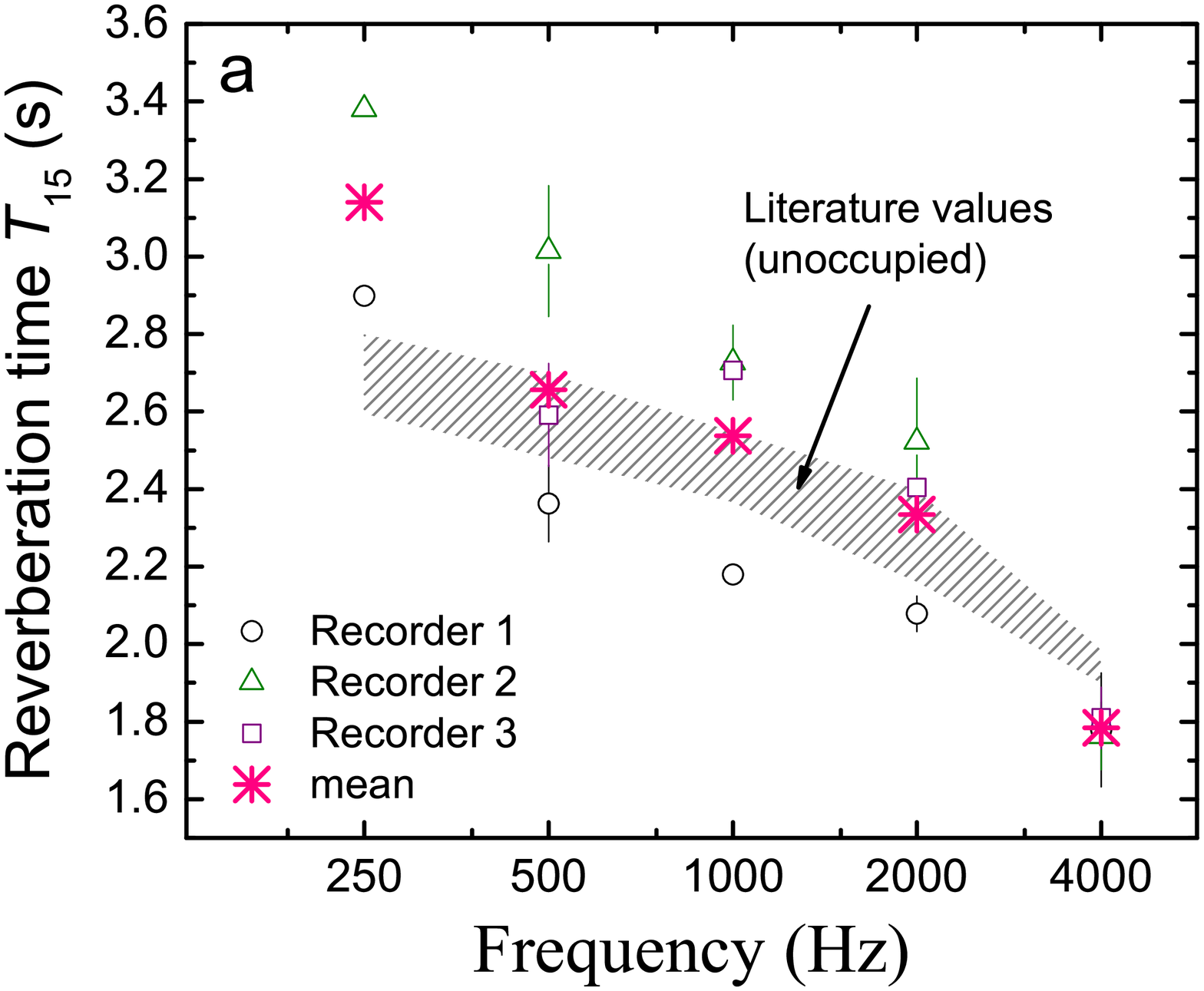} \\
		\includegraphics[width=0.8\columnwidth]{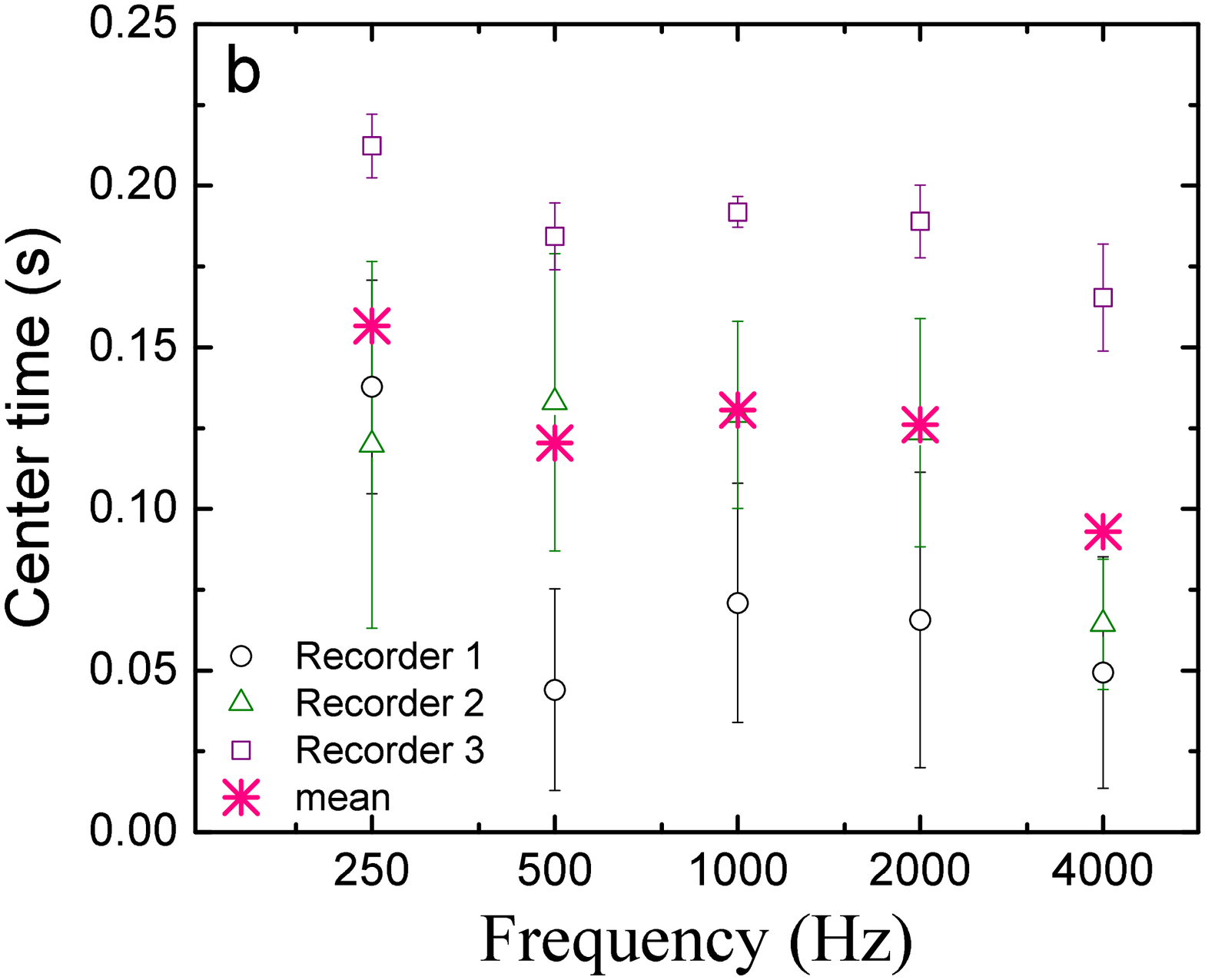} 
    \end{center}
    \caption{(a) Reverberation time of the auditorium for different octave bands. The error bars correspond to the standard deviation from five hand-claps. The asteroids are mean values from the different recorders. The region shaded with gray lines corresponds to the literature values \cite{Garai2015, Mueller2007}. (b) Corresponding center time at different frequency.}
    \label{fig:rt}
\end{figure}

The reverberation time obtained from different hand-claps and recorders is shown in Fig.\,\ref{fig:rt}(a). A comparison between the mean value obtained from the current measurement and the literature values \cite{Garai2015,Mueller2007} demonstrates that RT can be extracted from the simple HCSP approach, at least in a certain frequency range.  However, due to the lack of high repeatability, different hand-claps yield slightly different RTs (represented as errors). Thus, an average of the obtained RTs from different clapping events and recorders is necessary for a sensible characterization. The error from the exponential fits can be ignored as it is typically much smaller than the uncertainty from different clapping events. 

The deviation from the literature values at low frequencies ($\le 250$ Hz) can be attributed to the following three reasons: (i) The sound energy does not always decay to below $-20$dB within the recording time period because of the slow decay rate for low frequency sound. Therefore, the statistics is not sufficient for an accurate quantification. (ii) The strong influence of the direct sound may lead to large fit error because the exponential decay may start at an energy level smaller than the standard $-5$~dB. (iii) The frequency response of MEMS (microelectromechanical systems) microphones, which are typically used in smart phones, has a roll-off at low frequencies and a peak at $\sim 15$kHz due to the design of the chamber geometry \cite{Michaelis2017}. Consequently, the reliability of data obtained at low frequencies may suffer from the low SNR. 

RT obtained at different locations is compared in Table~\ref{tab}. Quantitative agreement between the data from the auditorium and from the stage is found for frequency $\ge 500$~Hz, suggesting that the characterization of RT is weakly dependent on where the hand-claps and recorders are located. Due to the same reason described above, there is also an overestimation of RT at $250$~Hz for the RT obtained on the stage.

\begin{table}
\caption{\label{tab}%
Mean values of the reverberation time obtained in the auditorium (A) and on the stage (S), unoccupied.}
\begin{ruledtabular}
\begin{tabular}{c|ccccc|c}
\textrm{}&
\textrm{250}&
\textrm{500}&
\textrm{1000}&
\textrm{2000}&
\textrm{4000}&
\textrm{(Hz)} \\
\colrule
A & 3.1 & 2.7 & 2.5 & 2.3 & 1.8 & (s) \\
S & 3.5 & 2.6 & 2.5 & 2.1 & 1.8 & (s) \\
\end{tabular}
\end{ruledtabular}
\end{table}

Figure~\ref{fig:rt}(b) shows the center time obtained from the same measurement. It is defined as the first moment of the squared impulse response $t_{\rm s} = \int_0^{\infty}[p(t)^2]t{\rm d}t/\int_0^{\infty}[p(t)^2]{\rm d}t$ \cite{Kuttruff2000}. As $t_{\rm s}$ characterizes the balance of direct and reflected sound, it is expected to vary with the locations of the recorders and sound sources, as well as with the source signals generated. Therefore, the data scattering is strong in comparison to the RT results. As described above, the influence of direct sound on recorder 1 is the largest among the three, therefore the corresponding $t_{\rm s}$ obtained is the smallest for most frequencies. An average over the results from different frequencies yields $0.125\pm0.023$~s, which also compares fairly well with a previous measurement \cite{Garai2015}. This value suggests that the speech intelligibility is $\ge80$\% \cite{Kuttruff2000} if the speaker is standing close to the proscenium. 

\section*{Conclusions and Outlook}
\label{sec:sum} 

To summarize, room acoustic characterizations are conducted for the Bayreuth Festspielhaus using hand-claps as sound source and smart phones as recording devices. The reverberation time obtained with the ITA-Toolbox agrees quantitatively with another measurement \cite{Garai2015} taken place in the same month within a certain frequency range. Possible reasons for the data scattering among different recorders and hand-claps, as well as deviations from the literature values are discussed. This investigation demonstrates the possibility of using amateur measurement devices for monitoring room acoustics, provided that the following precautions are properly taken care of:

\begin{itemize}
    \item[-] The sound generated by hand-claps should be as loud as possible to have sufficient initial sound energy, otherwise the accuracy for individual measurements will suffer from the low SNR.
    \item[-] Sufficient waiting time (at least $1.5t_{\rm s}$ with $t_{\rm s}$ the expected RT) between individual claps is necessary for extracting the response at low frequencies, because the low frequency components decay slower than the high frequency ones.
    \item[-] Multiple measurements (hand-claps, recording devices and locations) are needed for better statistics.
\end{itemize}

In addition to RT, the center time obtained with the same recordings shows strong scattering because of its dependence on the positions of the sound source and recording devices. Therefore, further analysis on the spatial distribution of $t_{\rm s}$ is needed for a better comparison with the other measurements. 

The possibility of quantifying room acoustics properties with easily accessible devices helps, for instance, to monitor an opera house in occupation on a more regular basis and to provide instantaneous feedback on building open-air theaters with the help of real-time analysis tools \cite{Seetharaman2012, Rosenkranz2017}. 

\section*{Acknowledgments}

I acknowledge the organizers of the 19th Dynamics Days Europe for arranging the visit to the Bayreuth Festspielhaus. I would also like to thank Erik Werner, Thomas M\"uller, and Stephan Messlinger 
for their kind help in audio recording.

\end{document}